\documentstyle[epsfig,subfigure]{mn}
\newcommand{\cthead}[1]{\multicolumn{1}{c}{#1}}
\newcommand{\ks}{km~s$^{-1}$}
\newcommand{\kss}{km~s$^{-1}$ }

\renewcommand{\thesubfigure}{\alph{subfigure}}
\makeatletter
\renewcommand{\@thesubfigure}{(\thesubfigure)\space}
\makeatother

\title[Space-VLBI observations]{Space-VLBI observations of OH maser OH34.26$+$0.15: \\
low interstellar scattering}

\author[V.I. Slysh et al.]
{V.I. Slysh,$^1$
 M.A. Voronkov,$^1$
 V. Migenes,$^2$
 K.M. Shibata,$^3$
 T. Umemoto,$^3$
 \newauthor
 V.I. Altunin,$^4$
 I.E. Val'tts,$^1$
 B.Z. Kanevsky,$^1$
 M.V. Popov,$^1$
 A.V. Kovalenko,$^1$
 \newauthor
 E.B. Fomalont,$^5$
 B.A. Poperechenko,$^6$
 Yu.N. Gorshenkov,$^6$
 B.R. Carlson,$^7$
 \newauthor
 S.M. Dougherty,$^7$
 J.E. Reynolds,$^8$
 D.R. Jiang,$^9$
 A.I. Smirnov$^1$ and
 \newauthor
 V.G. Grachev$^{10}$\\
   $^1$Astro Space Centre of Lebedev Physical Institute, Profsoyuznaya 
84/32, 117810 Moscow, Russia \\
   $^2$University of Guanajuato, Department of Astronomy, Apdo
Postal 144, Guanajuato, CP36000, GTO, Mexico \\
   $^3$National Astronomical Observatory, 2-21-1 Osawa, Mitaka, Tokyo 181,
Japan\\
   $^4$Jet Propulsion Laboratory, 4800 Oak Grove Dr., Pasadena, CA 91109,
USA\\
   $^5$National Radio Astronomy Observatory, 520 Edgemont Rd.,
Charlottesville, VA 22903, USA\\
   $^6$Special Research Bureau, Moscow Power Engineering Institute,
Krasnokazarmennaya st. 14, 111250 Moscow, Russia\\
   $^7$Dominion Radio Astrophysical Observatory, Herzberg Institute of
Astrophysics, National Research Council, PO Box 248, \\
\hphantom{$^7$}Penticton, BC, Canada V2A 6K3\\
   $^8$Australia Telescope National Facility, PO Box 76, Epping, NSW 2121,
Australia\\
   $^9$Shanghai Astronomical Observatory, 80 Nandan Rd, Shanghai 200080,
China\\
   $^{10}$Institute of Applied Astronomy, Zhdanovskaya str. 8, 197042
St.Petersburg, Russia}

\date{Received date; accepted date}
\pagerange{\pageref{firstpage}--\pageref{lastpage}}
\pubyear{1999}

\def\LaTeX{L\kern-.36em\raise.3ex\hbox{a}\kern-.15em
    T\kern-.1667em\lower.7ex\hbox{E}\kern-.125emX}

\begin{document}
\setcounter{dbltopnumber}{3}

\label{firstpage}

\maketitle
\begin{abstract}
We report on the first space-VLBI observations of the OH34.26$+$0.15
maser in two main line OH transitions at 1665 and 1667 MHz. The
observations involved the space radiotelescope on board the
Japanese satellite {\it HALCA} and an array of ground radio telescopes.
The map of the maser region and images of individual maser spots were
produced
with an angular resolution of 1 milliarcsec which is several times higher
than
the angular resolution available on the ground. The maser spots were
only partly resolved and a lower limit to the brightness temperature
$6\times10^{12}$~K was obtained. The maser seems to be located in the
direction of low interstellar scattering, an order of magnitude lower
than the scattering of a nearby extragalactic source and pulsar.
\end{abstract}

\section{Introduction}
OH masers were the first to be discovered as a new class of astronomical
phenomenona exhibiting intense, narrow band, polarized, variable
emission in interstellar molecular transitions~\cite{Weaver}. The source
of this emission was shown to be small, compact clumps of neutral gas, at
the periphery of compact \hbox{H\,{\sc ii}} regions created by newly born
massive O-stars. Another class of OH masers is related to evoled low-
and medium-mass stars, and is not considered here. High angular resolution
study of OH masers showed that they consist of a number of bright compact
maser spots separated by several arcseconds. The spots themselves are from
2 to 70 milliarcseconds in extent, and in some masers are barely resolved,
even with the highest angular resolution available from ground based
VLBI arrays~\cite{Moran}.

The goal of the high angular resolution study
of OH masers is to determine relative position of the spots, to measure the
intensity, and to map the
spots. Determination of the size and the shape of the maser spots would
be very helpful for understanding the maser emission mechanism and its
limiting brightness temperature. The shape of the maser spots may be
indicative of the type of physical phenomena responsible for the origin
of the maser emission.

Several possible sites for the origin
of maser emission;
such as the shocks at the border of \hbox{H\,{\sc ii}}
regions, or the interaction region between a molecular outflow and
ambient molecular clouds, as well as protoplanetary accretion disks around
young star, have been suggested.
The maser spot shape and size are important properties of the maser
emission, provided they are
intrinsic to the source, and not caused by a propagation effect such as
interstellar scattering.

There is a widespread opinion that OH masers
are heavily scattered by interstellar turbulence, and that observed
images of maser spots are scattering broadened images of essentially
point-like sources. This view is based on data obtained from the study
of scattering of pulsars and continuum sources. The importance of
scattering for OH masers is due to their location within the thin
galactic disk where the scattering is most severe. The first observational
evidence for the scattering of OH masers obtained by direct VLBI measurements
came from Kemball, Diamond and Mantovani~\shortcite{Kemball}. They measured
the size of the strongest spectral features in 16 galactic masers using a
single-baseline interferometer with a fringe spacing of 5 milliarcsec.
Only three 
masers out of 16 observed have shown compact details, less than 2.5
milliarcsec in
size. The rest of the masers were completely resolved implying angular size
larger than 5 milliarcsec. Kemball et al.~\shortcite{Kemball} concluded that their
results were consistent with the interstellar scattering angle of about
20 milliarcsec derived from pulsar scattering data, and with the clumpy distribution
of electron density fluctuations which are responsible for the large spread
of measured angular sizes. Hansen et al.~\shortcite{Hansen} made a VLBI
survey of interstellar broadening of 20 galactic OH masers, using a
7-station VLBI array. They found a large range of angular sizes from
0.8$\pm$0.5~milliarcsec to 74$\pm$7~milliarcsec in their sample of OH masers. They fitted
the angular size $\theta$ versus distance $D$ dependence by a
relation
$\theta\sim D^{0.65}$
which is close to the expected scaling with distance for homogeneous
turbulence with a Kolmogorov power law spectrum. They concluded that the
measured sizes result from interstellar scattering which is strongly
non-uniform.

Among the OH masers measured in these VLBI experiments,
several are unresolved to the limit of angular resolution.
Slysh et al.~\shortcite{Slysh} used a 3-station VLBI network to measure the
angular size of three such compact OH masers. With a fringe separation of
4.2~milliarcsec and high signal to noise ratio data the angular size of several
spectral features in the two main line OH transitions were measured
simultaneously. The
measured angular size was in the range from 1.4$\pm$0.4~milliarcsec to
4.3$\pm$0.1~milliarcsec. It was found that the angular size of the spectral
features with the same radial velocity, presumably originating in the same
region, was larger in the 1667~MHz line than in the 1665~MHz line. This cannot
be the result of scattering, since the scattering size should be virtually the 
same at these two close frequencies (the scaling between 1665~MHz and 
1667~MHz is negligible!).
It was concluded that the measured
angular size at 1667~MHz 4.3$\pm$0.1~milliarcsec for OH34.26$+$0.15 and
3.5$\pm$0.2~milliarcsec for W48 is intrinsic to the sources. These two masers
together with OH45.47+0.13 exhibit an angular size for
spectral features which is much less than the average broadening estimated
from pulsar measurements. The conclusion of that study was that the
distribution of the scattering material in the Galaxy is patchy, and that
these three masers are located in a direction of low interstellar
scattering. For this reason these masers were selected for space-VLBI
measurements using the Japanese satellite {\it HALCA} and a ground telescope
array.
In this paper we report the first results of the space-VLBI observations
of OH maser OH34.26$+$0.15.

\section{Observations and data reduction}

\begin{figure*}
\subfigure[ground array (AT, BL, Mop, Sh, Tid, Usu)]{
\label{uvGround}
\resizebox{0.46\textwidth}{0.46\textwidth}{\includegraphics{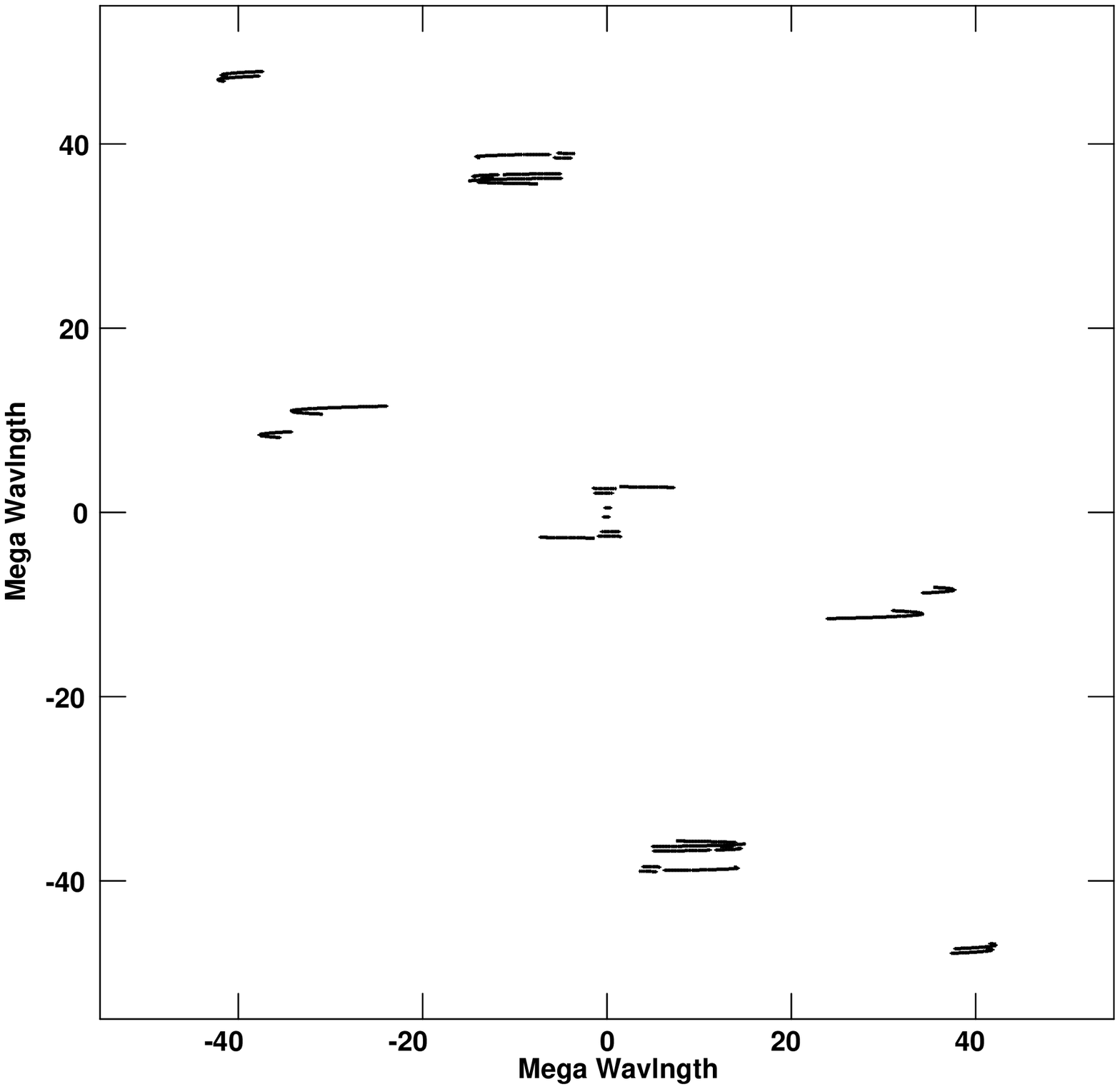}}
}\hfill
\subfigure[ground array + {\it HALCA}]{
\label{uvVSOP}
\resizebox{0.46\textwidth}{0.46\textwidth}{\includegraphics{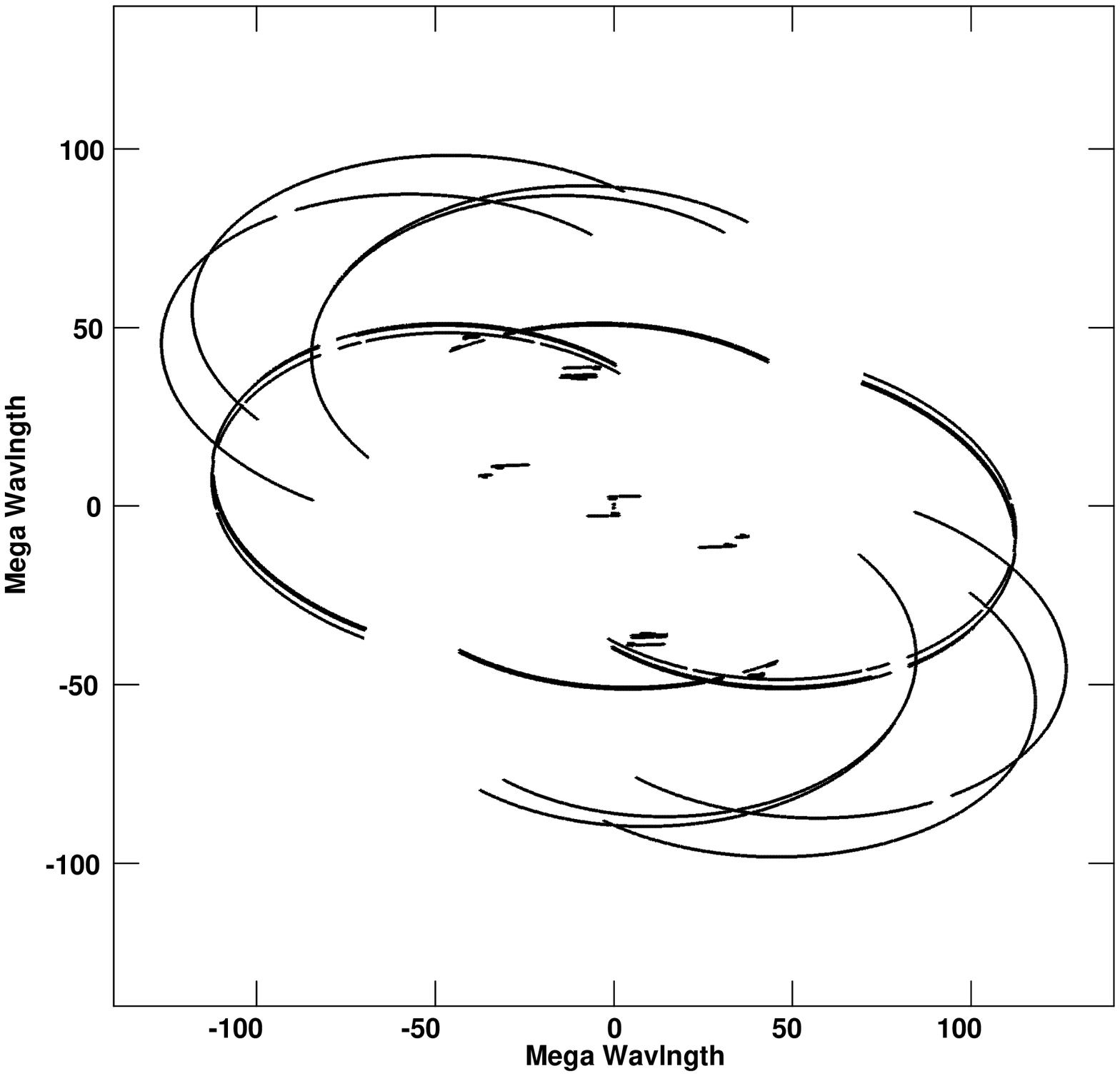}}
}
\caption{uv-plane coverage for OH34.26$+$0.15. Note different scale
for (a) and (b).}
\label{uvplots}

\end{figure*}

The maser source OH34.26$+$0.15 was observed as a part of the
Key Science Program of the
Japanese Satellite {\it HALCA} on March 24-25, 1998, by the space-ground
very long baseline interferometer. The satellite radio
telescope has an 8-m diameter deployable parabolic mirror and uncooled L-band
receiver, orbiting the Earth with a 6 hour period, apogee 21000~km and
perigee 560~km~\cite{Hirabayashi}. The ground radio telescope array in this
particular session included the 70-m DSN telescope Tidbinbilla in Australia
(Tid),
the phased 6$\times$22m Australia Telescope (AT), the 22-m Mopra Telescope in
Australia (Mop),
the Shanghai 25-m Telescope in China (Sh),
the 64-m Telescope in Usuda Japan (Usu),
and the 
64-m Telescope in Bear Lakes (BL) near Moscow (Russia). It was the first time
that the Bear Lakes Telescope participated in space-VLBI observations.
Table~\ref{Antennas} is a list of telescopes with their parameters:
system noise temperature T$_{sys}$, gain and rms noise in {\it HALCA} to
ground telescope baselines after 5 minutes of integration.
The uv-coverage\- of the ground telescope array is shown on
Fig.~\ref{uvGround}, and the uv-coverage\- of the full array including the
space radio telescope is shown on Fig.~\ref{uvVSOP}.
It is evident that the addition of the space radio telescope not only
increases the uv-coverage by a factor of two due to the high orbit of the
space radio telescope, but also significantly fills the
uv-plane\- coverage at lower resolution due to the orbital motion
of satellite.
For the ground array the
synthesized beam was 5.8$\times$2.1~milliarcsec, and for the space-ground array
the beam was 1.8$\times$1.0~milliarcsec.

Left circular polarization data over a 16 MHz
bandwidth centered on the 1665~MHz and 1667~MHz OH main lines
were recorded
using S2 recorders at each station. The data were correlated using the
Canadian S2 Space-VLBI correlator\footnote{The Canadian S2 Space-VLBI
correlator at the National Research Centre of Canada, Dominion Radio
Astrophysical Observatory is operated by the Canadian Space
Agency.} \cite{Carlson}. The 16 MHz data were filtered using a FIR
bandpass filter and then re-sampled before correlation in order to
attain a spectral resolution of 488 Hz~channel$^{-1}$\-  (0.088~\ks)
over a bandpass of 500 kHz.
Amplitude calibration was achieved using the ``gain'' and
T$_{sys}$ data (Table~\ref{Antennas}) obtained from ground observatories
and from the VSOP operations group. The post-correlation data reduction was
performed using the {\sc aips} package of NRAO. 
\begin{table}
\caption{List of the telescopes used in the experiment.}
\label{Antennas}
\begin{tabular}{lcrr}
\cthead{Station}&\cthead{T$_{sys}$}&\cthead{Gain}&\cthead{rms}\\
&\cthead{K}&\cthead{K Jy$^{-1}$}&\cthead{Jy}\\
\hline
{\it HALCA} \hfill (satellite) & 88 & 0.0043 & ------\\
Australia Telescope~~~~ \hfill (Australia)    & 40 & 0.667 & 0.58\\
Bear Lakes \hfill (Russia)    & 72 & 0.54 & 0.41\\
Mopra \hfill (Australia)   & 40 & 0.1 & 0.61\\
Shanghai  \hfill (China) & 86 & 0.060 & 1.35\\
Tidbinbilla  \hfill (Australia) & 41 & 0.9 & 0.51\\
Usuda  \hfill (Japan) & 70 & 0.797 & 0.46\\
\hline
\end{tabular}
\end{table}
\begin{figure*}
\parbox{0.46\textwidth}{
\resizebox{0.46\textwidth}{0.46\textwidth}{\includegraphics{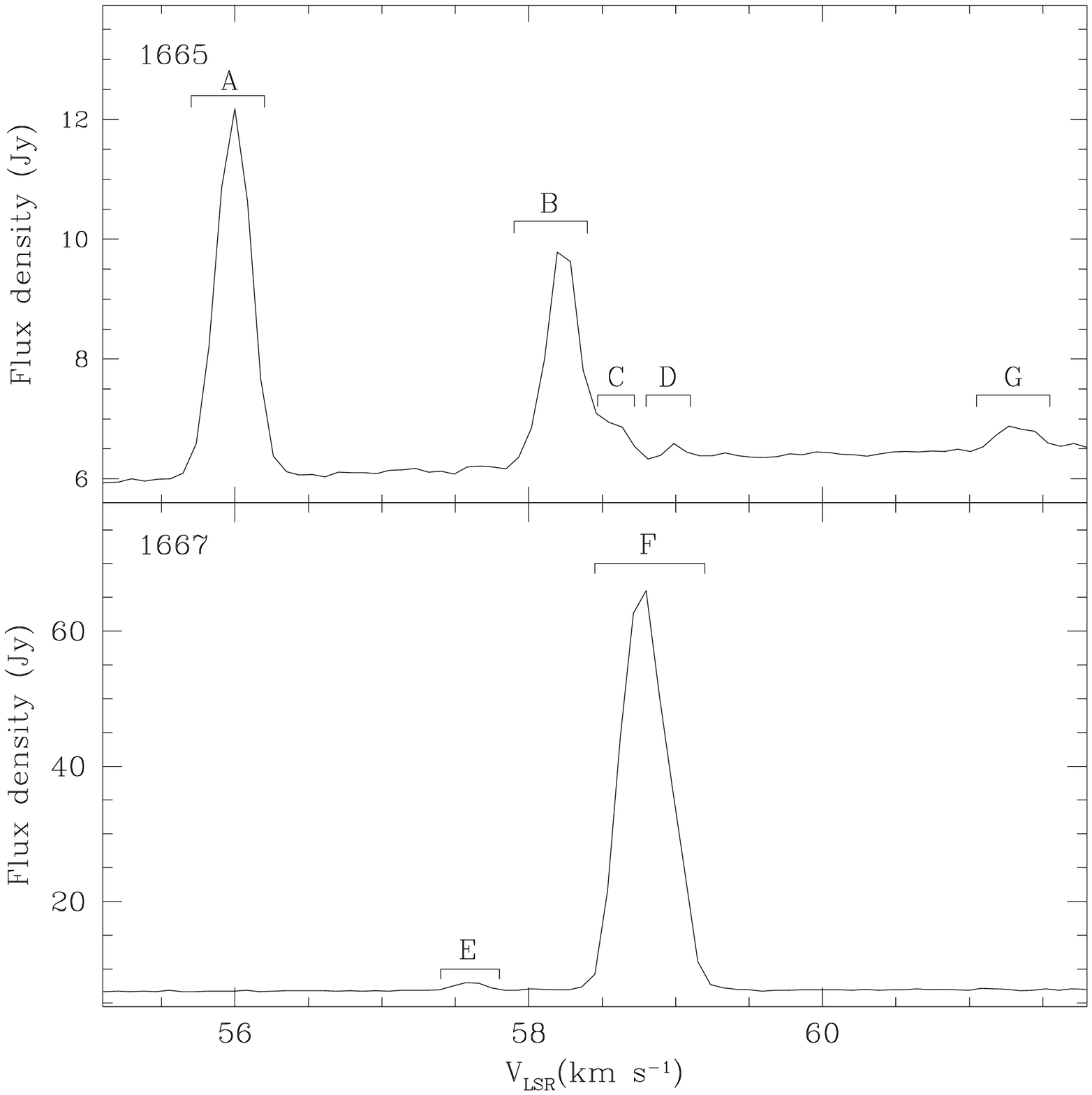}}
\caption{Correlated spectrum of OH34.26$+$0.15 on the short base line
AT--Mopra. Upper panel: the 1665~MHz line; lower panel: the 1667~MHz line.}
\label{Spectra}}
\hfill
\parbox{0.46\textwidth}{
\resizebox{0.46\textwidth}{0.46\textwidth}{\includegraphics{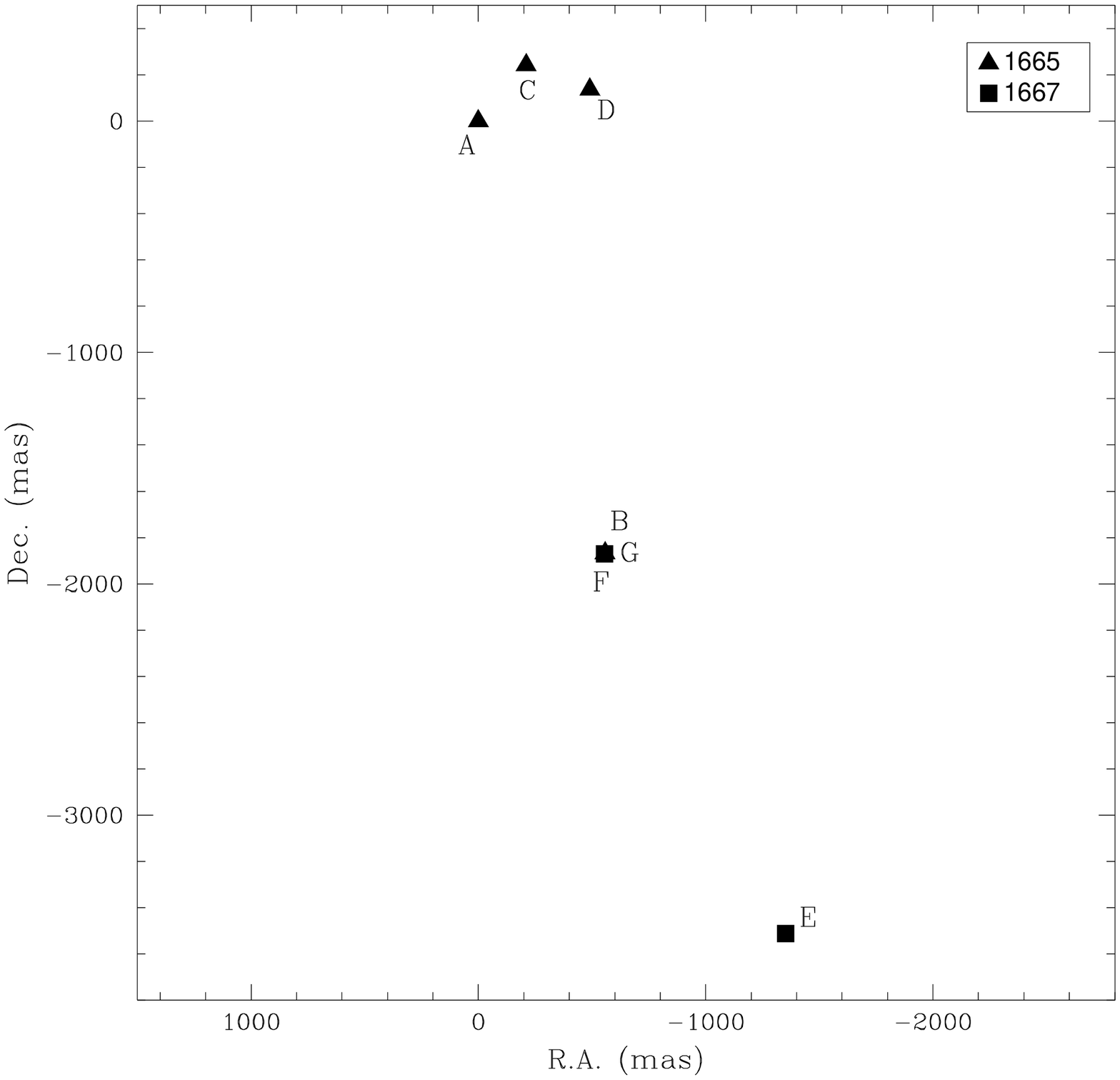}}
\caption{Map of OH34.26+0.15. Triangles denote 1665~MHz features,
squares are 1667~MHz features. Feature A was taken as a reference (0,0).}
\label{Position}}
\end{figure*}

\begin{figure*}
\subfigure[ground array map. Contours are 0.971$\times$(1, 2, 3, 4, 5, 6,
7, 8, 9)~Jy/beam]{
\label{mapGround}
\resizebox{0.46\textwidth}{0.46\textwidth}{\includegraphics{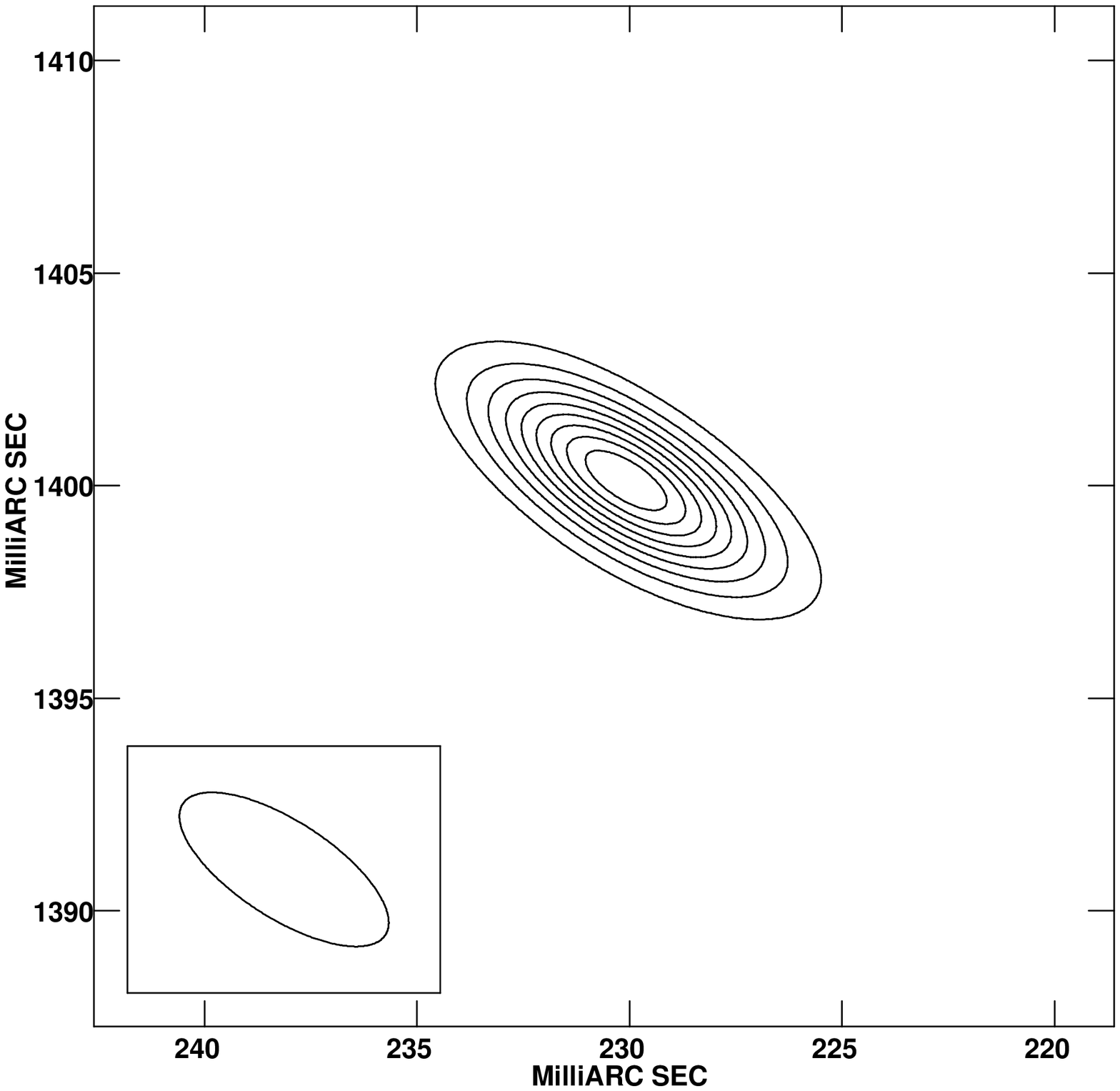}}
}\hfill
\subfigure[ground array + {\it HALCA}. Contours are 0.577$\times$(1, 2, 3, 4, 5,
6, 7, 8, 9)~Jy/beam]{
\label{mapVSOP}
\resizebox{0.46\textwidth}{0.46\textwidth}{\includegraphics{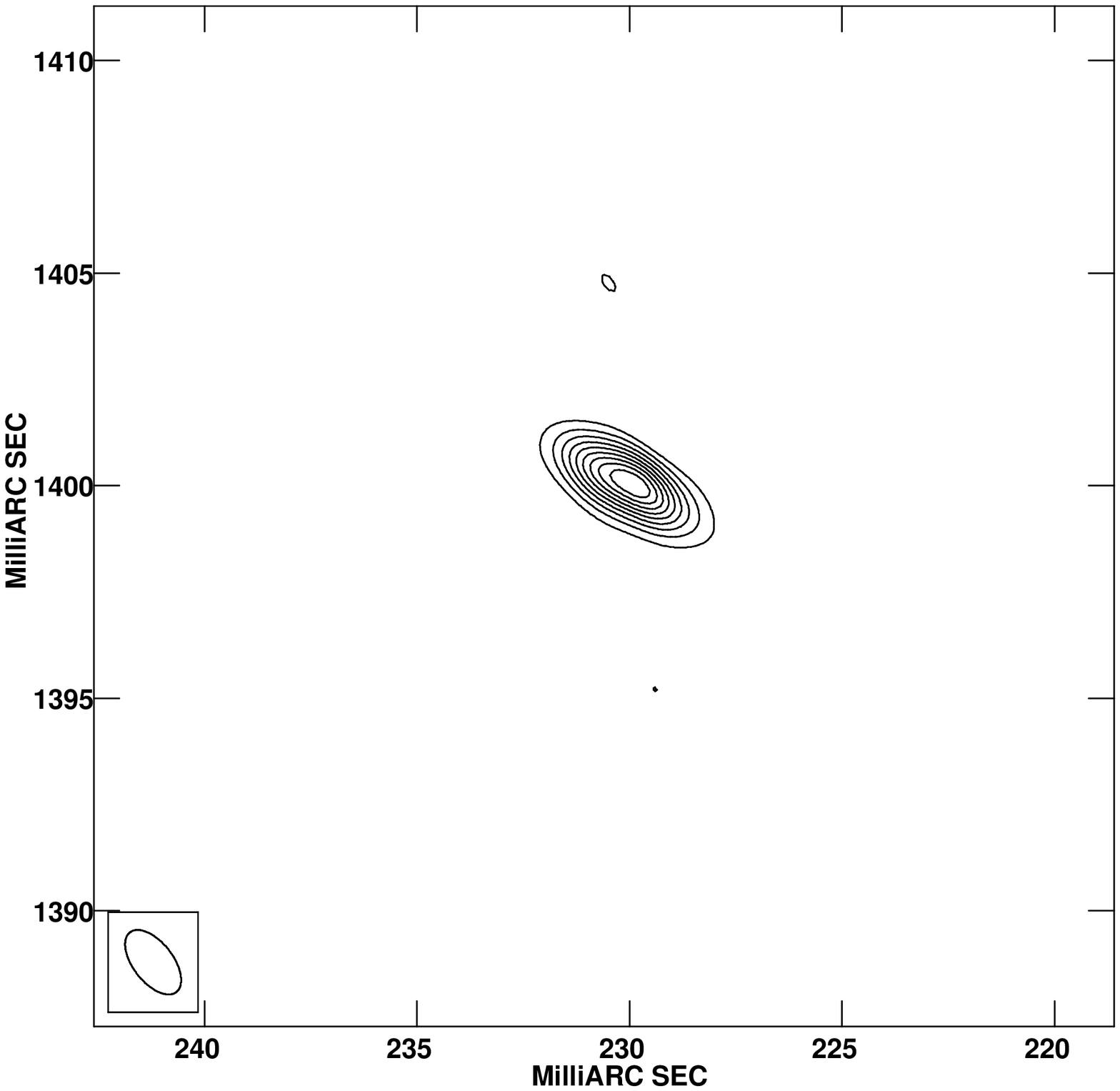}}
}
\caption{Map of the 1665~MHz feature A at a radial velocity 56.0~\ks.
The synthesized beam is shown in the lower left corner. The offsets are
relative to the correlating position: $\alpha=18^h53^m18\fs67$,
$\delta=01\degr14\arcmin58\farcs5$ (J2000.0)}
\label{A_Maps}
\end{figure*}

Phase jitter was discovered at the Bear Lakes telescope which caused
splitting of the fringe spectrum into several components.
It was found that the phase of the Bear Lakes telescope receiver experienced
quasi-periodical excursions with a period of about 200 seconds. The phase
jitter was removed by means of phase self-calibration of the
ground array with a short solution interval of 30 seconds, which is less
than the instrumental, tropospheric and ionospheric phase variation 
period. The multiple spectral components 
disappeared, and the central peak increased by a factor of two.

The strongest and most compact
spectral feature at the radial velocity 56~\kss
in the 1665~MHz spectrum was taken as a reference in the self-calibration
process. Its position relative to the correlating position
was determined by the fringe rate
method, using the ground array. The measured offset is 0\farcs23 in right
ascension and 1\farcs4 in declination, with an accuracy of
0\farcs1, and our best position for this feature is
$\alpha=18^h53^m18\fs685$, $\delta=01\degr14\arcmin59\farcs9$ (J2000.0).
The position of the reference feature A determined here differs from the
measured with the VLA by Gaume \& Mutel~\shortcite{GaumeMutel} by
about 0\farcs15 in both coordinates, which is within the combined error
limits of the two measurements. After applying this correction the residual
fringe rate an all base lines including those to {\it HALCA} was  found
to be close to zero, and the phase variations became slow enough to
allow integration time up to 5 minutes.

The delay errors for the ground
array telescopes were measured to be less than 300 nanoseconds from
observation of the continuum calibration source 1730$-$130, which was
observed during the time interval when {\it HALCA} was not observing.
The maximum residual delay associated with the baselines to {\it HALCA}
obtained during correlation of the experiment using the Penticton 
Correlator, which assumed an accurate {\it HALCA} orbit and other known
apriori information, was less than 300 nanoseconds at both 1665~MHz and
1667~MHz. Thus, any additional delay errors after correlation are 
unlikely to be larger than 300 nanoseconds which will produce a 
negligable phase difference over the 5.3~\kss spread across maser lines.

The relative positions of the 1665~MHz line maser spots with respect to
feature A were
determined in two steps.
First, an approximate position relative to the reference feature was
measured by the fringe rate method with the ground array. These measurements
provided positions with an accuracy of about 50 milliarcseconds, with
relative displacements of the features reaching 2000 milliarcseconds.
Then maps of the separate features were constructed centered on these
approximate positions, with the ground only or space-ground array when
the signal to noise ratio was large enough. The features mapped by the
space-ground array are marked by stars in Table~\ref{Features}. The
imaging and cleaning were done with the {\sc aips} task {\sc ``imagr''},
and the features were analyzed with the {\sc aips} task {\sc ``sad''} from 
which the flux density, position and deconvolved size (or size limit) 
were obtained for each feature. The relative positional accuracy is
about 1 milliarcsecond. 
Result of this analysis is given in
Table~\ref{Features}, where for each Gaussian component are given:
radial velocity, relative position, flux density, angular size of the
major and minor axis of deconvolved Gaussian components as well as position
angle of the major axis. 

The 1667~MHz OH line data were reduced in the following manner.
The 58.70~\kss feature in F is the strongest feature and that was
used as the phase reference. Images of the other features in F and 
feature E (after using fringe-rate mapping to determine its 
approximate position), were then made by imaging and cleaning as for
the 1665~MHz line. In order to tie the 1665~MHz and 1667~MHz images to
one coordinate system, we assumed that feature E at 1667~MHz was at
the position ($-$1350,$-$3530) milliarcseconds displaced from feature A
at 1665~MHz, following Gaume \& Mutel~\shortcite{GaumeMutel}. This 
registration should be accurate to about 50 milliarcseconds. The near 
coincidence of 1665~MHz features B and G with that of F at 1667~MHz 
is independent evidence that this registration is accurate. 

\section{Results}
\begin{table}
\caption{Spectral features}
\label{Features}
\begin{tabular}{llrrrrrr}
&\cthead{Velocity}&\cthead{$\Delta\alpha$}&\cthead{$\Delta\delta$}&\cthead{Flux}
&\multicolumn{2}{c}{Size}&\\
&&&&\cthead{density}&\cthead{maj}&\cthead{min}&\cthead{PA}\\
&\cthead{km s$^{-1}$}&\multicolumn{2}{c}{milliarcsec}&\cthead{Jy}&
\multicolumn{2}{c}{milliarcsec}&\cthead{\degr}\\
\hline
\multicolumn{8}{c}{1665 MHz}\\
A&56.0 & 0 & 0 & 10.5 &2.0 &$<$0.5& 70\\     
B&58.15 & -553 & -1873 & 0.6 & 3.6 & 2.7 & 88\\
&58.15 & -557 & -1867 & 2.9 & 5.9 & 2.4 & 156\\
&58.19 & -553 & -1869 & 0.6 & $<$2.7 & $<$2.7 & -----\\    
&58.25 & -528 & -1864 & 0.6 & 3.1 & 1.7 & 33\\
C&58.63 & -210 &  242  & 2.1 & 3.1 & 0.2 & 70\\
D&58.98 & -491 &  137  & 1.7  &$<$1.8&$<$1.8& -----\\
G&61.30& -553 & -1866 & 1.0  &1.6 & 0.6 & 88 \\
\multicolumn{8}{c}{1667 MHz}\\
E&57.63 & -1350 & -3530 & 4.5 & 5.0 & 2.9 & 150\\
F&58.65 & -550 & -1890 & 19.7 & 6.8 & 2.8 & 155\\
&58.70 & -553 & -1888 & 21.3 & $<$2.5 & $<$2.5& -----\\
&58.75 & -553 & -1885 & 19.7 & 5.1 & 1.8 & 112\\
&58.95 & -549 & -1836 & 19 & 6.5 & 4.1 & 170\\
\hline
\end{tabular}
\vskip 3mm
\begin{description}
\item[1] -- reference position of A is  $\alpha$=18$^h$53$^m$18\fs685;
$\delta$=01\degr14\arcmin59\farcs9 (J2000.0)
\item[2] -- Reference position of E for 1667~MHz is ($-$1350,$-$3530) 
milliarcseconds with respect to A. 
\end{description}
\end{table}
\begin{figure*}
\subfigure[Map of the 1665~MHz~~ feature B.
Contours are 0.10$\times$(1, 2, 3, 4, 5, 6, 7, 8, 9)~Jy/beam]{
\label{mapcompB}
\resizebox{0.46\textwidth}{0.46\textwidth}{\includegraphics{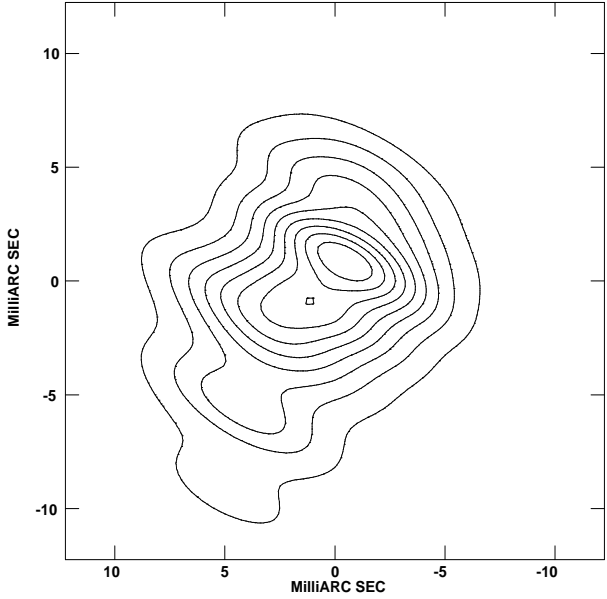}}
}\hfill
\subfigure[Map of the 1667~MHz~~ feature F.
Contours are 2$\times$(1, 2, 3, 4, 5, 6, 7, 8, 9)~Jy/beam]{
\label{mapcompF}
\resizebox{0.46\textwidth}{0.46\textwidth}{\includegraphics{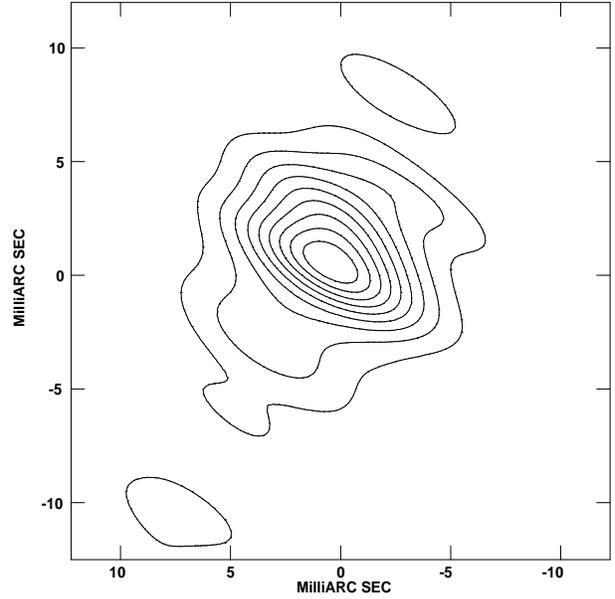}}
}
\caption{Maps of the 1665~MHz feature B and 1667~MHz feature F.}
\label{Comps}
\end{figure*}

The correlated LCP spectra of OH34.26+0.15 on the short Australia
Telescope -- Mopra baseline on which the source is unresolved, in both main line OH
transitions is shown in Fig.~\ref{Spectra}. Five spectral features A--D
and G
are present in the 1665~MHz spectrum, and two features E and F in the 1667~MHz
spectrum; their relative position is shown in Fig.~\ref{Position}.
Fig.~\ref{mapGround}
shows a map of feature A obtained with the ground array. The feature
is unresolved with a synthesized beam of 5.8$\times$2.1~milliarcsec (lower left
corner). Fig.~\ref{mapVSOP} shows a map of the same feature A obtained
with the space-ground array having a synthesized beam of 1.8$\times$1.0~milliarcsec
(lower left corner). The feature is partly resolved, and this is evident
from the different position angles of the beam and the image.
A deconvolved fitted Gaussian to this image is an ellipse
at a position angle of 70\degr, with the major
axis 2.0~milliarcsec and the minor axis equal to or less than 0.5~milliarcsec.
Inspection of Table~\ref{Features} shows that all
components are partly resolved with the beam of the space-ground or ground
only array, but most of the components may have structure and consist of
several smaller components.

Other 1665~MHz spectral features are offset from the reference feature
by up to 1\farcs9, their position
is shown in Fig.~\ref{Position}. Feature B consists of four separate
closely spaced spots as can be seen on the map (Fig.~\ref{mapcompB}) and
in Table~\ref{Features}, where radial
velocity, relative position, flux, and angular size of all spectral
features are given. Feature F (1667~MHz) coincides with feature B within
the measurement errors, and its shape is similar to the shape of
feature B (Fig.~\ref{Comps}).
These two features may come from the same maser condensation, a
supposition that can be
tested with higher accuracy relative 1665/1667 position measurements.
If this is the case the radial velocities of features B and F must coincide,
and the
observed velocity difference of 0.5~\kss can be attributed to a difference
between Zeeman splittings of 1665~MHz and 1667~MHz lines because of
different g-factors of two transitions. If the observed features B and F are
$\sigma$-components of Zeeman pairs (both are left circular polarized), one
can calculate that the velocity difference of 0.5~\kss corresponds to
the magnetic field of B=4.2~milligauss, which is a value commonly found in
OH masers~\cite{Garcia_Barreto}.

\section{Discussion}
\subsection{Maser spot distribution}

OH34.26$+$0.15 is located at a distance of 3.8~kpc~\cite{Reid}
near a cometary
\hbox{H\,{\sc ii}} region and near two ultra-compact {H\,{\sc ii}} regions.
Maser spots A, C and D coincide with the northern ultra-compact
\hbox{H\,{\sc ii}} region G34.3$+$0.2B, while spots B, E, F and G
coincide with the cometary \hbox{H\,{\sc ii}} region
G34.3+0.2C (see Fig.~7 in Gaume, Fey and Claussen~1994). Our map of OH
components (Fig.~\ref{Position}) is consistent with the VLA map of the
same source obtained by Gaume \& Mutel~\shortcite{GaumeMutel} in 1985.
All components A--E have their counterparts on the Gaume \& Mutel~
\shortcite{GaumeMutel} map, except component D. On the other hand, some
weaker components from the map of Gaume \& Mutel~\shortcite{GaumeMutel}
are not present in our map. The splitting of components B and F into four
subcomponents is not evident on the Gaume \& Mutel~\shortcite{GaumeMutel}
map because of the much lower resolution of the VLA. The relative position
of the components on both maps is the same within the VLA position error
of about 50~milliarcsec, while the relative positional errors in
our map are about 1~milliarcsec. The mean
difference in position is 5.5$\pm$25~milliarcsec. 
\begin{figure}
\resizebox{\hsize}{!}{\includegraphics{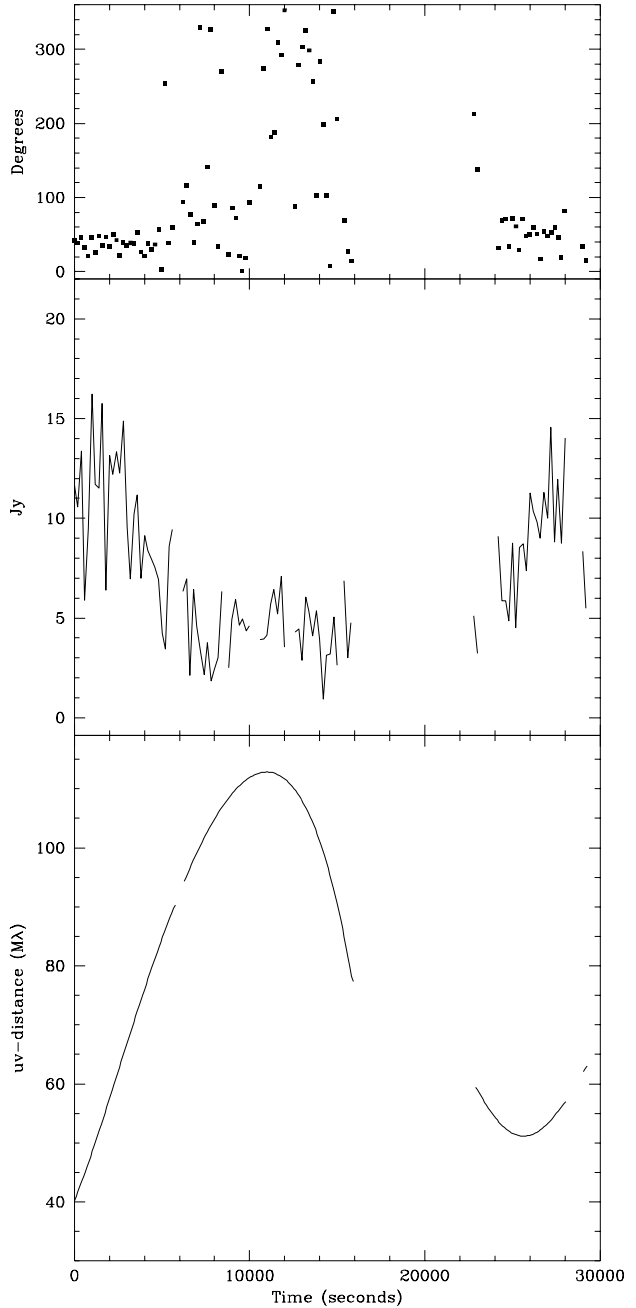}}
\caption{Fringe amplitude and phase versus time for feature A on
the baseline AT--{\it HALCA}. Upper panel -- phase, middle panel -- amplitude,
lower panel -- uv-distance.}
\label{Fringes}
\end{figure}

\subsection{Resolution of the feature A}
\label{Resolution_A}

The most compact component is feature A, which
is known to be left hand circularly polarized~\cite{Caswell83}.
It was
resolved in one direction, and the angular size in this direction is
2.0~milliarcsec and less than 0.5~milliarcsec in the perpendicular direction.
Feature A was fitted with an ellipse  in position angle 70\degr$\,$ with
the ratio of major to minor axis greater than 4. This size is a result of the
fitting and deconvolution with the beam of the image shown in
Fig.~\ref{mapVSOP}. As an almost independent check of the size of
feature A we made a direct fit of a Gaussian model to uv-data using only
fringe amplitude information. This procedure is independent in the sense
that it does not use the self-calibration and global fringe fitting.
The best fit to fringe amplitudes from three telescopes AT, Tid and
{\it HALCA} was achieved with an elliptical Gaussian, having major and minor
axis 2.6$\pm$0.5 and 0.3$\pm$0.3~milliarcsec and a position angle 61\fdg5$\pm$3\fdg2
which, within the errors, is the same as obtained
from the image. From our data it is not possible to say whether maser
spot A has a stripe-like shape, or is composed of a chain of
point-like sources. To determine this, one needs higher angular resolution
and higher signal to noise data in order to obtain maps with a higher
dynamic range.

In this experiment we could not use the full angular
resolution of the space-ground array because of the limitations set by
the signal to noise ratio of the data.
While the orbit of {\it HALCA} provided
uv-distances up to 135~M$\lambda$, the fringes were lost in the noise 
every time the uv-distance from a ground
telescope to the space telescope exceeded about 90~M$\lambda$.
This can be seen in Fig.~\ref{Fringes}, where we
show fringe amplitude and phase versus time for the baseline AT--{\it HALCA}.
During the session the uv-distance first increased from 40~M$\lambda$ and
reached a maximum of 113~M$\lambda$ at 11000 seconds, then decreased to
78~M$\lambda$ at 16000 seconds. It is evident from Fig.~\ref{Fringes} that
the fringe amplitude has decreased from about 12~Jy to less than 4~Jy
which is below the noise level. This can be seen from the phase behaviour in the
upper panel of Fig.~\ref{Fringes}: the phase varies smoothly from 0 to
about 6000 seconds (about 90~M$\lambda$), and then becomes chaotic.
The smooth phase variation resumes during the second portion beginning
at about 23000 seconds, up to the end of this session. During the second
portion the uv-distance was between 51 and 64~M$\lambda$ and the fringe
amplitude was above the noise. This example shows that the low signal to noise
ratio prevents us from using the full angular resolution available. If
the signal to noise ratio were increased by a factor of 3 it would be
possible to implement full angular resolution imaging. The main cause
of the low signal to noise ratio is the small size of the space radio
telescope. An increase in the signal to noise ratio by a factor of three
would require an increase in the space telescope diameter by
the same factor, or a combination of larger diameter and lower system
noise temperature.

The upper limit of the angular size of feature A cited above corresponds
to the lower limit of its brightness temperature of 6$\times10^{12}$~K. Other
features have somewhat larger sizes and lower brightness temperature limits,
although it cannot be excluded that they consist of several unresolved
subcomponents. This value exeeds the brightness temperature predicted by
some OH maser models. For example a model by Pavlakis \&
Kylafis~\shortcite{Pavlakis} predicts brightness temperatures about
$10^{11}$~K, or at least an order of magnitude lower than observed. On the
other hand  there is an absolute upper bound to the brightness temperature
imposed by the stimulated transition line broadening (or resonant Stark
effect~\cite{Slysh73}). The stimulated emission rate $R$ is
\begin{equation}
\label{StRate}
R=\frac\Omega{4\pi}\frac{kT_B}{h\nu}A=889\frac\Omega{4\pi}\times T_{B12}\;\;{\rm s}^{-1}
\end{equation}
where $\frac\Omega{4\pi}$ is maser emission solid angle, $A$ is the spontaneous
emission rate ($7.11\times10^{-11}$~s$^{-1}$ for 1665~MHz OH transition), and
$T_{B12}$ is the brightness temperature in units of 10$^{12}$~K. For
feature A the stimulated emission rate
$R=\frac\Omega{4\pi}5.4\times10^3$~s$^{-1}$, and is comparable to the
observed  line width which is 0.33~\ks, or 1.6$\times10^3$~Hz. For an isotropic
maser $\Omega=4\pi$, and the contribution of the stimulated emission to the
line width is $\frac R{2\pi}=8.5\times10^2$~Hz, or about a half of the
linewidth. This means that the brightness temperature can not be much higher
than the observed upper limit $6\times10^{12}$~K, unless the maser emission
is highly directed. For example, if $\frac\Omega{4\pi}=10^{-2}$, the
brightness temperature can be as high as $6\times10^{14}$~K. Since the maser
lines do not show any measurable broadening  the maser is probably
anisotropic.

Elongated shapes of OH masers features are also seen in other sources.
Baudry \& Diamond~\shortcite{Baudry} found a filamentary and arc-like structure in the OH maser
W3(OH) at 13.44~GHz. They relate this structure to a shocked environment
of the ultra-compact \hbox{H\,{\sc ii}} region. One can also speculate that the
elongated filaments were formed as a result of an anisotropic compression
of the molecular gas into dense condensations in the presence of the
magnetic field which is observed in OH masers through Zeeman splitting
and polarization of emission lines.

\subsection{Interstellar scattering}

OH masers connected with regions of formation of massive stars are located
in the plane of the Galaxy, and are subject to strong interstellar
scattering. As a result of scattering, the image of small sources is
broadened to an extent that depends on the level of turbulence that
produces scattering electron density inhomogeneities. The observed small
angular size of maser spots in OH34.26$+$0.15 puts a strong upper limit on
the turbulence level which is usually characterized by a constant $C_N^2$
in the turbulent electron density power spectrum. The scatter-broadened
angular size is related to the so-called scattering measure
\begin{equation}
\label{SM}
SM=\int\limits_0^LC_N^2\;ds\mbox{,}
\end{equation}
where the integral is over distance $s$ up to the source at distance $L$.
For galactic sources the scattered angular diameter $\theta_s$ and $SM$ are
related by the expression~\cite{Taylor}
\begin{equation}
\label{ThetaS}
SM=\left(\frac{\theta_s}{71~mas}\right)^{\frac 53}\nu_{GHz}^{\frac{11}3}
~kpc~meters^{-\frac{20}3}\mbox{.}
\end{equation}
If the measured upper limit of the size of the minor axis of component
A of 0.5~milliarcsec is due to scattering, and $\nu=1.6654$~GHz, then from
equation~(\ref{ThetaS}) one has
\begin{equation}
\label{SMfinal}
SM=1.68\times10^{-3}~kpc~meters^{-\frac{20}3}\mbox{.}
\end{equation}
At a distance of 3.8~kpc this gives
$C_N^2=4.4\times10^{-4} meters^{-\frac{20}3}$. This is almost an order of
magnitude less than the average found from pulsar scattering measurements,
where 
$C_N^2=10^{-2.5}~m^{-\frac{20}3}$ (fig.~13 in Cordes, Weisberg \& Boriakoff
1985). This means that scattering towards OH34.26$+$0.15 is much less
than the average in the Galaxy.

There exist measurements for interstellar
scattering from a nearby extragalactic source 1849+005, which is only 0\fdg76
from OH34.26$+$0.15~\cite{Fey}. For this source the angular size measured
at the frequency 1465~MHz is 560~milliarcsec. Scattering was found to 
scale with frequency as
$\nu^{-1.65}$ in the range from 333~MHz to 4860~MHz. From pulse broadening,
Frail \& Clifton~\shortcite{Frail} estimated the angular size of a
highly scattered pulsar PSR~1849$+$00 (which is within 10\arcmin$\,$
 of 1849$+$005)
as 500~milliarcsec at 1~GHz. If both values are scaled to the OH-line frequency of
1665~MHz, one
has 450~milliarcsec from 1849+005 and 216~milliarcsec from the pulsar. These are two to
three orders of magnitude larger than the size of the maser spots in
OH34.26$+$0.15. We note that 1849$+$005 is an extragalactic source, and
PSR 1849$+$00 is a very distant pulsar at a distance of 14.5~kpc~\cite{Frail}.
Therefore, the ray path from the sources to the Sun traverses large distances
in the
Galaxy, with the closest approach to the Galactic Centre of about 4 kpc. 
The maser OH34.26$+$0.15 is on the same path, but is much closer to the Sun,
at 3.8~kpc. If most of the scattering material lies beyond that distance,
presumably near the point where the path is closest to the Galactic Centre
(at 6.6~kpc from the Sun, assuming the distance to the Galactic Centre from
the Sun 8 kpc), then the OH maser emission will not be scattered
so much.
One could argue that the large discrepancy of scattering parameters
between OH maser and continuum sources (like pulsars) is due to small
scale variations
in the distribution of the scattering material. The angular separation
between the OH maser and 1849$+$005 is 0\fdg76 and corresponds to 50~pc. This
small scale is inconsistent with an almost equally small scattering of two
other OH masers in this region OH35.20$-$1.76 and OH45.47$+$0.13~\cite{Slysh}
which are separated by about 10\degr, or 650~pc. Based on observations of
three low scattered OH masers in the same region, we suggest that
there is a large scale deviation in the distribution of the scattering
material in the Galaxy. 

\section{Summary and conclusions}

We report on the first space-VLBI observations of OH masers. It was also
the first time that the 64-m telescope at Bear Lakes near Moscow participated
in a space-VLBI experiment. We constructed a map of the maser OH34.26$+$0.15
with angular resolution better than 1~milliarcsec. The maser consists of small
bright spots distributed over several arcseconds, the size of the spots
in some cases is less than 0.5~milliarcsec, and the brightness temperature is
larger than 6$\times10^{12}$~K. This OH maser source experiences
an exceptionally
small interstellar scattering, and we propose that there is a strong
large scale variation in the distribution of the scattering material in the
Galaxy.

\section{Acknowledgements}

We gratefully acknowledge the VSOP project, which is led by the Japanese
Institute of Space and Astronomical Science in cooperation with many
organizations and radio telescopes around the world. The National Radio
Astronomy Observatory is a facility of the National Science Foundation,
operated under a cooperative agreement by Associated Universities, Inc.

\label{lastpage}
\end{document}